%% file: main.tex
\documentclass[a4paper, conference]{IEEEtran}
\IEEEoverridecommandlockouts
\usepackage{cite}
\usepackage{amsmath,amssymb,amsfonts}
\usepackage{algorithmic}
\usepackage{graphicx}
\usepackage{textcomp}
\usepackage{xcolor}
\usepackage{svg}

\usepackage{esvect}
\usepackage{tikz}
\usepackage{pgfplots}
\usepackage{booktabs}
\usepackage{multirow}
\usepackage{bbm}
\usepackage{bbold}
\usepackage{dsfont}
\usepackage{url}
\usepackage[a4paper, total={180mm,250mm}]{geometry}

\def\BibTeX{{\rm B\kern-.05em{\sc i\kern-.025em b}\kern-.08em
    T\kern-.1667em\lower.7ex\hbox{E}\kern-.125emX}}
\pgfplotsset{compat=1.18}

\usepackage{fancyhdr}
\fancypagestyle{firstpage}
{
    \fancyhead[L]{\footnotesize \textcopyright 2026 IEEE.  Personal use of this material is permitted. Permission from IEEE must be obtained for all other uses, in any current or future media, including reprinting/republishing this material for advertising or promotional purposes, creating new collective works, for resale or redistribution to servers or lists, or reuse of any copyrighted component of this work in other works. The paper is accepted at IEEE DFTS'26.}
    \fancyhead[R]{}
}

\begin{document}

\title{CheckOne: Lightweight Fault Detection and Mitigation for Vision Transformers \\
\thanks{This paper is supported in part by  EU Grant Project 101160182 “TAICHIP”, and the EU Grant 101194287 “NexTArc”.}\vspace{-3mm}}






\author{
\IEEEauthorblockN{
Mohammad Hasan Ahmadilivani\IEEEauthorrefmark{1},
Sven-Markus Loorits\IEEEauthorrefmark{1},
and Jaan Raik\IEEEauthorrefmark{1}}
\IEEEauthorblockA{\IEEEauthorrefmark{1}\textit{Tallinn University of Technology}, Tallinn, Estonia \\
\{mohammad.ahmadilivani, sven.loorits, jaan.raik\}@taltech.ee}\vspace*{-10mm}}

\maketitle

\thispagestyle{firstpage}

\begin{abstract}
The wide adoption of Vision Transformers (ViTs) in safety-critical applications raises reliability concerns related to hardware faults. Algorithm-Based Fault Tolerance (ABFT) methods have emerged as lightweight and symmetric protection mechanisms for DNNs. However, they are particularly challenging for ViTs due to their significant computational requirements. This work comprehensively evaluates the reliability of ViTs, emphasizing the need for symmetric protection in their layers. Furthermore, we present \textit{CheckOne}, a novel, cost-effective method for fault detection and mitigation in ViTs that significantly reduces the computational cost compared to conventional ABFT. Through extensive experiments with multiple ViTs, \textit{CheckOne} mitigates critical faults by up to $26\times$ and achieves an average $3.8\times$ higher performance than ABFT in ViTs. 

\end{abstract}


\input{sections/1-introduction}
\input{sections/3-method}
\input{sections/4-results}

\input{sections/5-conclusions}

\bibliographystyle{IEEEtran}
\bibliography{ref.bib}
\end{document}

%% file: sections/1-introduction.tex
\section{Introduction} \label{sec:intro}

The rapid evolution of Deep Neural Networks (DNNs) continuously introduces both new opportunities and challenges. 
Transformer architectures have emerged as a central focus recently \cite{vaswani2017attention}. Their superior performance is appealing for edge-based, safety-critical applications \cite{mondal2026transformer}; however, reliability and efficiency remain significant challenges. The substantial computational and memory demands of these models \cite{tay2022efficient}, combined with the complexity of their reliability assessment and overheads associated with redundancy-based fault tolerance \cite{ahmadilivani2024systematic}, further complicate their practical edge deployment.



Recent studies have demonstrated that Vision Transformers (ViTs) are highly vulnerable to hardware-induced faults. Beam experiments have revealed that soft errors can severely degrade the inference accuracy of ViT models \cite{badia2025reliability,roquet2024cross}. Simulation-based Fault Injection (FI) further confirms that transient faults affecting memories can propagate through transformer architectures, leading to substantial output corruption \cite{liao2025analyzing,ahmadilivani2026LBR,ahmadilivani2026effective}. Although these studies collectively establish the susceptibility of ViTs to hardware faults, they provide limited insight into the relative criticality of individual ViT components and the practical implications of single bit-flips during deployment.

To address this gap, this work presents a comprehensive investigation into the fault criticality of ViTs, 
through an extensive layer-wise analysis. 
Our observations reveal that \textbf{vulnerability is relatively uniform across transformer layers and blocks}, indicating that selective protection strategies may offer limited effectiveness and motivating the need for symmetric fault tolerance mechanisms for reliable ViT deployment.

\begin{figure}[t!]
\resizebox{0.5\textwidth}{!}{%
\begin{tabular}{cc}
    \includegraphics[width=0.25\textwidth]{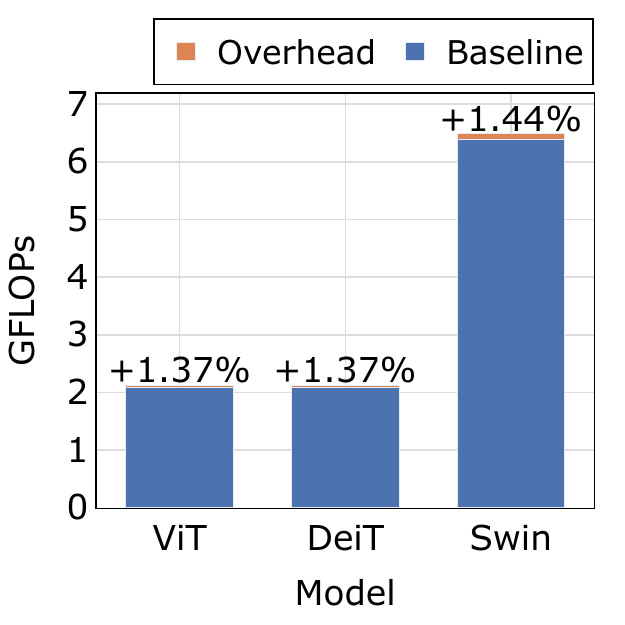}
 &  
    \includegraphics[width=0.25\textwidth]{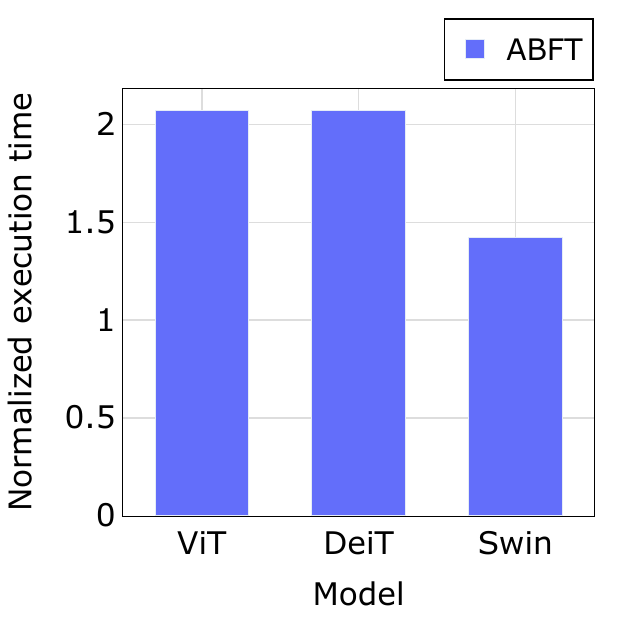}
\\
a) FLOPs overhead & b) Execution time overhead
\end{tabular}%
}
\caption{Overhead of conventional ABFT in ViT models on GPU. a) Theoretical operations overhead, b) execution overhead on GPU normalized to baseline ViTs execution.}
\label{fig:motivation}
\vspace{-7mm}
\end{figure}


Algorithm-Based Fault Tolerance (ABFT) techniques have emerged as lightweight and symmetric protection mechanisms for DNNs by augmenting matrix multiplication (GEMM) operations with checksum-based error detection and correction schemes \cite{zhao2020ft,ahmadilivani2026ftsparse}. These approaches typically introduce redundant checksum vectors for input and weight matrices and verify the integrity of the output through checksum recomputation during inference. Building upon this principle, several recent studies have proposed ABFT-oriented reliability mechanisms tailored to ViTs, incorporating architectural optimizations and transformer-specific adaptations to improve fault coverage and resilience \cite{liu2024alberta,ma2026error,liang2025attnchecker,dai2025ft,ma2023error,titopoulos2025custom}. Nevertheless, the exceptionally large matrix dimensions and intensive memory traffic characteristic of transformer architectures substantially limit the efficiency of existing ABFT implementations. In particular, checksum generation and verification for input, weight, and output matrices introduce significant memory-bound overheads, despite requiring relatively few additional arithmetic operations. As illustrated in Fig.~\ref{fig:motivation}, although the computational overhead remains below $1.5\%$, the resulting inference latency can increase dramatically, reaching up to $2.07\times$ longer execution time due to the dominance of memory-access costs during checksum processing.

To tackle this drawback, this work, for the first time, presents \textit{CheckOne}, an innovative method for on-the-fly calculation of checksums for efficient fault detection and removes the need for checksum computations for each matrix. It implicitly produces the summations through constant added vectors and compares the obtained summations with stored values in memory. The contributions of this work are as follows:

\begin{itemize}

    \item Presenting a comprehensive FI analysis of ViTs targeting both weights and activations to characterize the reliability of transformer blocks and linear layers. 

    \item Proposing \textit{CheckOne}, a novel and cost-efficient fault detection and mitigation mechanism for ViTs that reduces the critical SDC rate by up to $26\times$ compared to unprotected execution while 
    achieving an average $3.8\times$ performance improvement over conventional ABFT.

\end{itemize}


%% file: sections/3-method.tex
\section{Methodology} \label{sec:method}

\subsection{Fault Model}

With transistor scaling, 
SRAM-based on-chip memories become highly vulnerable to radiation-induced transient faults that manifest as bit-flips \cite{hill2021cmos}. 
In this work, we focus on single transient faults occurring in on-chip memories of DNN accelerators that affect parameter and activation matrices during inference. To model this behavior at the application level, a single bit-flip is randomly injected into either weights or input activations of linear layers. We assume that off-chip memory is protected by ECC mechanisms, correcting single-bit faults; consequently, pre-stored values remain error-free. 

\subsection{Reliability Evaluation} \label{subsec:fi}

We perform random Fault Injection (FI) campaigns targeting the input activations and weights of ViTs, for each individual layer, independently. During each inference, a single random bit is flipped in a random weight or input activation value represented in 32-bit floating-point. Resilience evaluation is performed using $5,000$ images from the ImageNet validation dataset. Each FI experiment is repeated $1,000$ times. To quantify resilience, we adopt two metrics: 1) average accuracy degradation, i.e., the difference between the baseline accuracy and the average accuracy across FI campaigns, and 2) critical SDC rate, i.e., the ratio of inference outputs whose predicted class differs from the corresponding fault-free classification. The experimental evaluation is conducted using three pre-trained ViT models on ImageNet: ViT-Tiny, DeiT-Tiny, and Swin-Tiny, all obtained from the \texttt{timm} library in PyTorch. Their baseline accuracies are $75.78\%$, $72.16\%$, and $81.48\%$, respectively. All experiments are implemented in PyTorch and executed on an NVIDIA A100 GPU.


\subsection{CheckOne: Lightweight Fault Detection and Mitigation}

The key idea in \textit{CheckOne} is to eliminate the need for recomputation in the output matrices after a GEMM operation in transformers and detect errors with minimal computational overhead. \textit{CheckOne} enables in-place computations of checksums by appending a vector of $1$s to the input rows and weight columns before a matrix multiplication in linear layers of a ViT, resulting in obtaining their summations by the GEMM computation. Thereafter, the \textit{CheckOne} method employs pre-computed weights and output activation range values to perform on-the-fly lightweight fault detection and localization, replacing errors with $0$.

\textit{CheckOne} consists of two main phases: 1) \textit{Offline Phase}, where the structure of a ViT is modified, and the detection values are obtained, and 2) \textit{Online Phase}, where the fault detection and mitigation are conducted during inference. Note that \textit{CheckOne} specifically targets single bit-flips in the inputs and weights of linear layers. 
In the \textit{Offline Phase}, the golden values for fault detection in the \textit{Online Phase} are obtained using validation data. For each linear layer, the weights and inputs are organized as a 2D matrix for a GEMM operation, and two golden sets are exported: 1) \textit{golden weight sums} $\vv{S^l_{w_{golden}}}$, i.e., the column-wise summation of the weights ($w$) for a linear layer $l$, 2) \textit{golden input range} $[Min(S^l_{x_{golden}}), Max(S^l_{x_{golden}})]$, i.e., the minimum and maximum input values ($X$) to a linear layer $l$. All values are stored and assumed to be fault-free during inference.

\begin{figure}[ht!]
    \centering
    \vspace{-4mm}
    \includegraphics[width=0.48\textwidth]{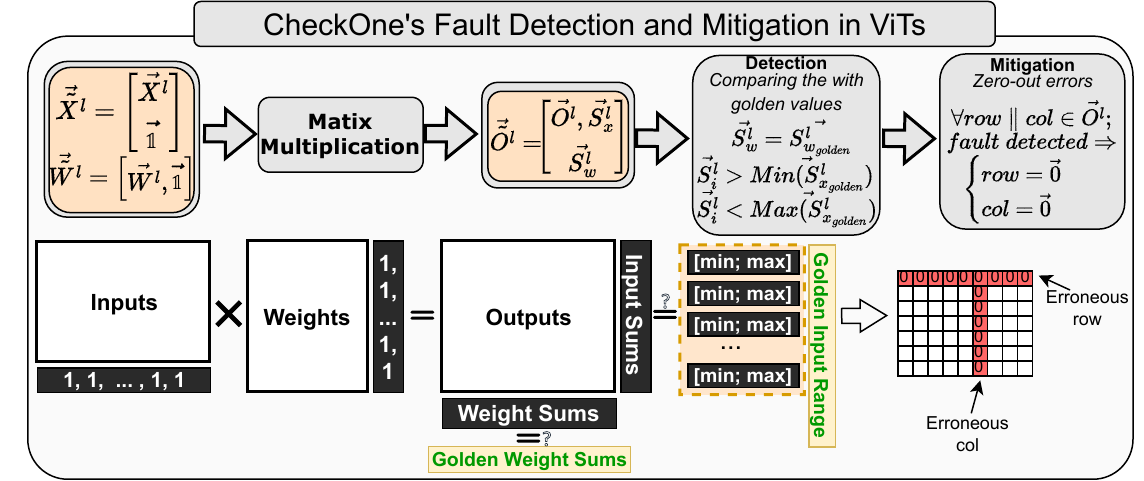}
    \vspace{-3mm}
    \caption{CheckOne method for each linear layer. }
    \label{fig:checkone}
    \vspace{-5mm}
\end{figure}

In the \textit{Online Phase}, the error detection and mitigation are conducted as shown in Fig. \ref{fig:checkone}. A linear layer $l$ performs a matrix multiplication between 2D arrays of input activations $\vv{X^l} \in \mathbb{R}^{m \times n}$ and weights $\vv{W^l} \in \mathbb{R}^{n \times k}$, producing output matrix $\vv{O^l} \in \mathbb{R}^{m \times k}$, while $m$, $n$, and $k$ represent the matrix dimensions. 

\vspace{-5mm}
\begin{equation}
    \vv{O^l} = \vv{X^l} \times \vv{W^l}
\end{equation}
\vspace{-5mm}

In \textit{CheckOne}, a vector $\vv{\mathbbm{1}}$ is appended to the rows of inputs and to the columns of weights, as shown in \ref{eq:checkone-mult}. The modified inputs $\vec{\tilde{X^l}} \in \mathbb{R}^{(m+1) \times n}$ and weights $\vec{\tilde{W^l}} \in \mathbb{R}^{n \times (k+1)}$ are multiplied, resulting in an output matrix $\vec{\tilde{O^l}}$, in which the last row contains the column-wise summation of weight matrix ($\vec{S^l_w}$), and the last column contains the row-wise summation of input activations ($\vec{S^l_x}$). The bottom right element of the matrix ($S_{\mathbbm{1}}$) contains the multiplication and accumulation of $\vv{\mathbbm{1}}$, which is equal to the dimension $n$. 

\vspace{-5mm}
\begin{equation}
    \vec{\tilde{O^l}} = \vec{\tilde{X^l}} \times \vec{\tilde{W^l}}  = \begin{bmatrix}\vec{X^l} \\ \vv{\mathbbm{1}} \end{bmatrix}  \times \begin{bmatrix}\vec{W^l}, \vv{\mathbbm{1}} \end{bmatrix} = \begin{bmatrix}\vec{O^l},  \vec{S^l_x} \\ \vec{S^l_w}, S_{\mathbbm{1}} \end{bmatrix}
    \label{eq:checkone-mult}
\end{equation}

To detect and mitigate faults in the input matrices of a GEMM operation at the output matrix $\vec{\tilde{O^l}}$, first, the vector $\vec{S^l_w}$ is compared element-wise with $\vv{S^l_{w_{golden}}}$. In the case of a mismatch, the corresponding obtained column in $\vec{\tilde{O^l}}$ is erroneous; thus, set to $\vv{\textbf{0}}$ to remove the effect of the faulty weight. Then, the vector $\vec{S^l_x}$ is compared with the range vector $[Min(S^l_{x_{golden}}), Max(S^l_{x_{golden}})]$. In the case that a value exceeds the range, the corresponding row in $\vec{\tilde{O^l}}$ is erroneous; thus, set to $\vv{\textbf{0}}$ to remove the effect of the faulty input activation.

\subsection{ABFT Implementation}


We implemented a conventional ABFT scheme for ViTs as a baseline for comparison. In this implementation, weight checksums are precomputed offline, stored in memory, and attached to the corresponding layers during inference, similarly to \textit{CheckOne}. In contrast, input and output checksums are generated dynamically at runtime. During inference, the output checksums are recomputed and compared against the expected checksum values derived from the stored weights. In the case of a mismatch, the erroneous output values and replaces with zero. 
In this approach, faults occurring in input activations cannot be detected.
Consequently, the reliability evaluation of the baseline ABFT method is restricted to FI campaigns targeting model weights and is compared with \textit{CheckOne}.

%% file: sections/4-results.tex
\section{Experiments} \label{sec:results}

\subsection{Reliability Evaluation}

\begin{figure*}[t!]
\centering
\begin{tabular}{cc}
    \includegraphics[width=0.45\textwidth]{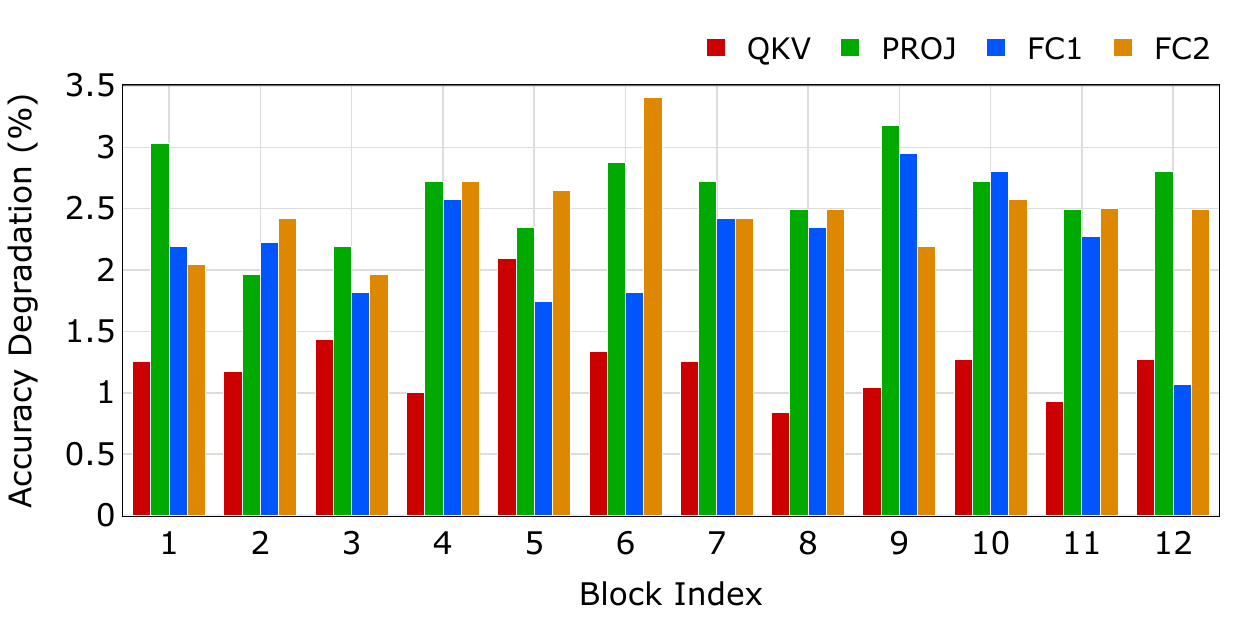}
 &  
    \includegraphics[width=0.45\textwidth]{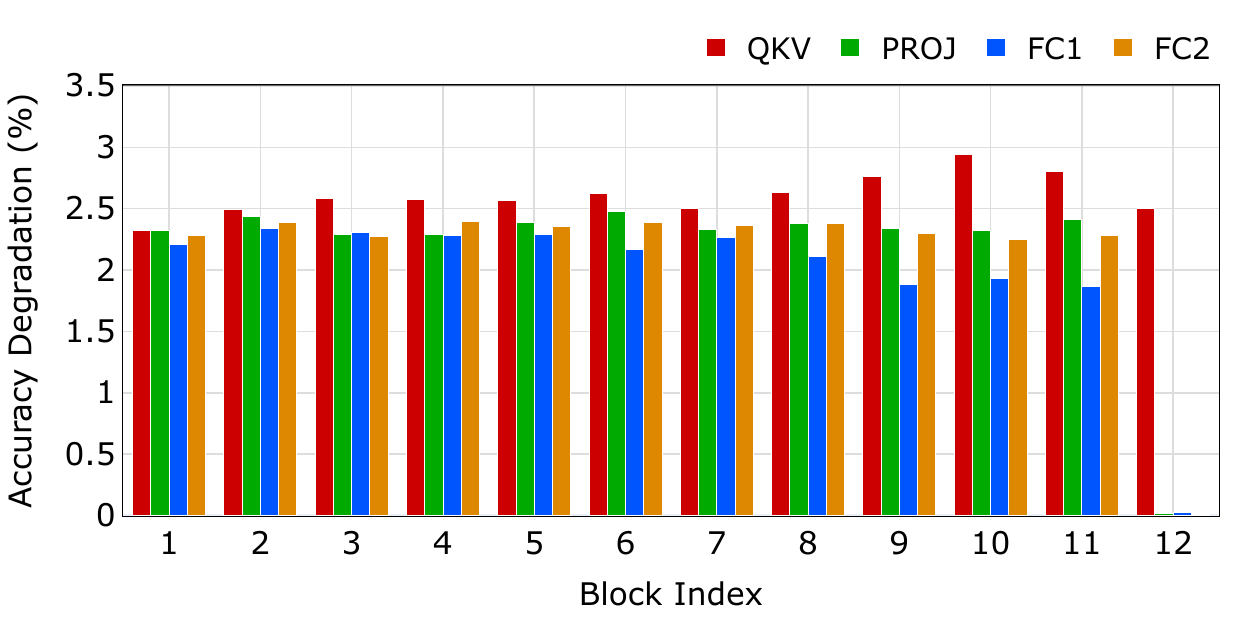}
\\
a) ViT-tiny FI into weight & b) ViT-tiny FI into input activations
 
 \\

    \includegraphics[width=0.45\textwidth]{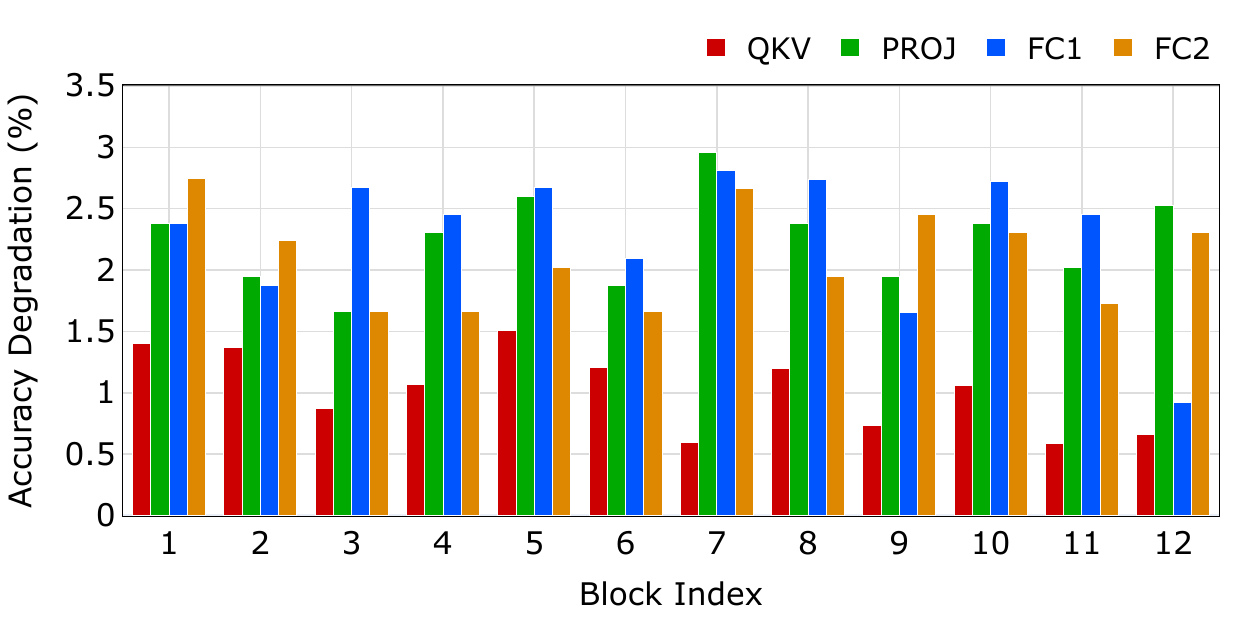}
 &  
    \includegraphics[width=0.45\textwidth]{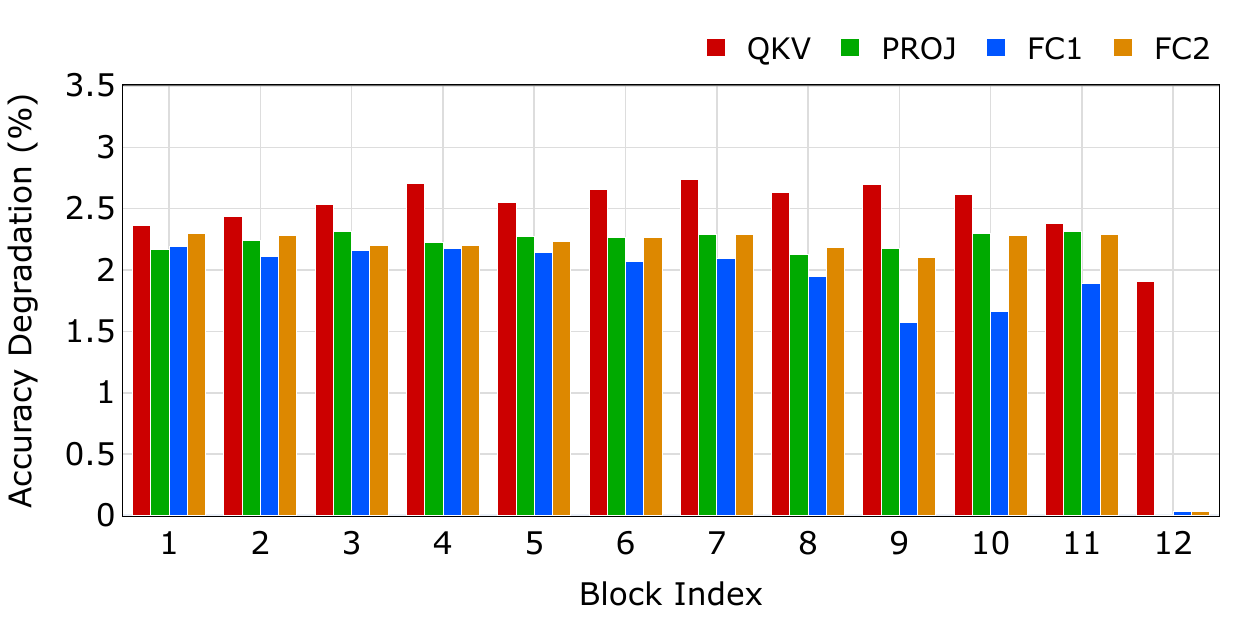}
\\
c) DeiT-tiny FI into weight & d) DeiT-tiny FI into input activations

 \\
    \includegraphics[width=0.45\textwidth]{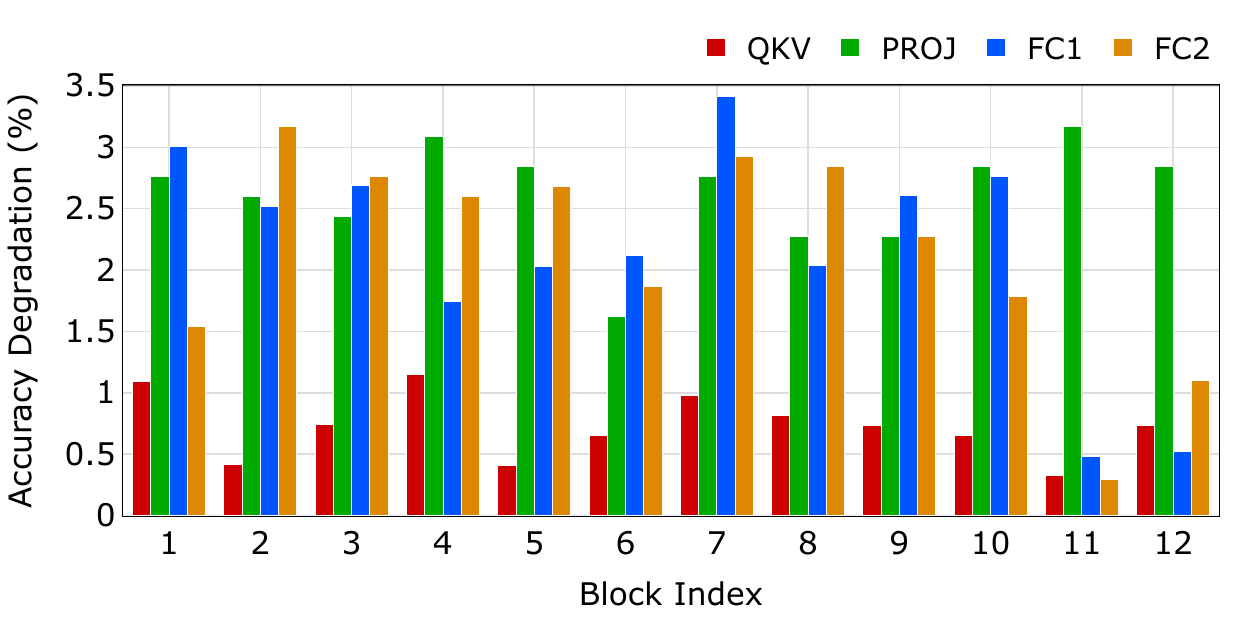}
 & 
    \includegraphics[width=0.45\textwidth]{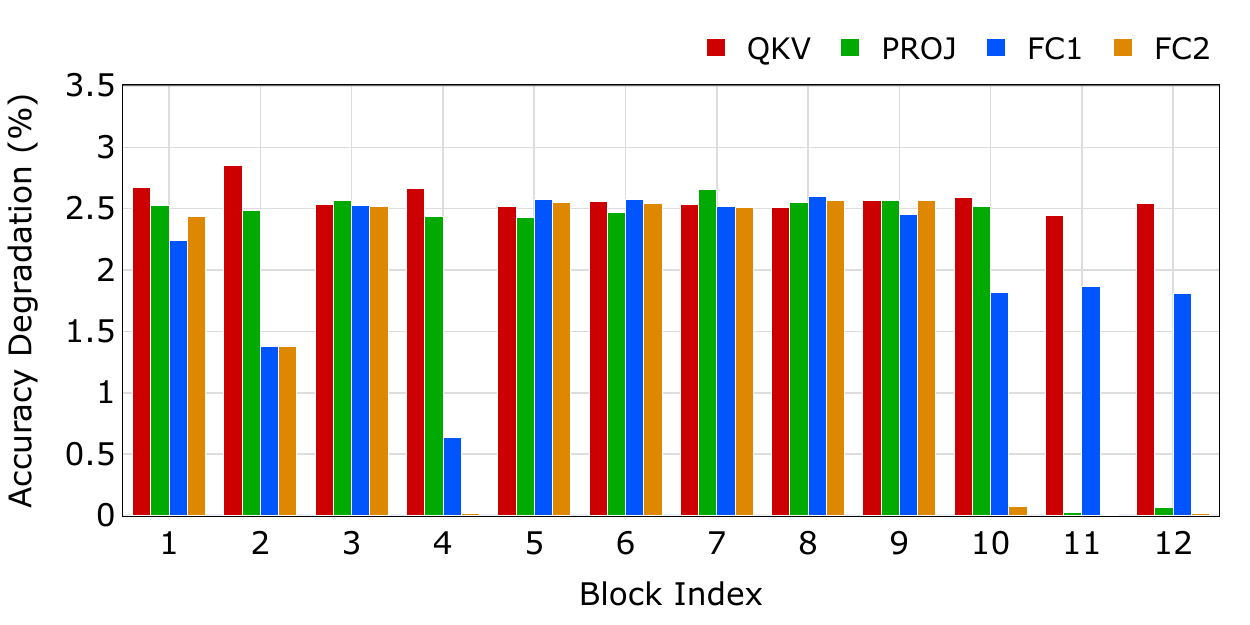}
\\
e) Swin-tiny FI into weight & f) Swin-tiny FI into input activations

\end{tabular}%
\caption{Accuracy degradation for ViT-tiny, DeiT-tiny, and Swin-tiny with layer-wise FI into weights and input activations.}
\label{fig:fi-results}
\vspace{-5mm}
\end{figure*}

Fig.~\ref{fig:fi-results} presents the average accuracy degradation obtained from layer-wise FI experiments. The results reveal that despite containing millions of parameters, ViTs are highly sensitive to even a single bit-flip in either weights or activations, leading to accuracy degradation of up to $3.41\%$. This observation emphasizes the necessity of effective fault protection mechanisms for transformer-based architectures. 

Across most transformer blocks, the QKV weights exhibit comparatively lower vulnerability, whereas the sensitivity of other layers varies significantly across models and blocks, preventing the identification of a consistent vulnerability pattern suitable for selective protection. In contrast, activation vulnerability remains relatively uniform across layers and blocks, although the final transformer blocks, particularly the FC2 layers in Swin-Tiny, demonstrate slightly higher resilience. 

Overall, considering both weight and activation FI, the vulnerability of ViT layers remains consistently high without a clearly dominant subset of critical layers. Moreover, layers exhibiting low sensitivity in activations often remain highly vulnerable in their weights, and vice versa. These findings indicate that \textbf{reliability enhancement techniques for ViTs should provide uniform and symmetric protection across all transformer layers and blocks} rather than relying on selective hardening strategies.

Table~\ref{tab:critical_sdc} reports the average critical SDC rate for each linear layer within the transformer blocks, averaged across all blocks. The results show that, although QKV weights are generally the least vulnerable parameters, the corresponding activations exhibit the highest vulnerability. Moreover, the critical SDC rates across different layers and models remain relatively close, indicating a uniformly high sensitivity to faults throughout the transformer architecture. These observations further confirm that effective reliability enhancement mechanisms for ViTs must provide symmetric protection across all layers, simultaneously covering both weights and activations.

        

\begin{table}[h!]
\vspace{-3mm}
    \centering
    \renewcommand{\arraystretch}{1.3}
    \caption{Average Critical SDC (\%) per linear layer, across all blocks of each unprotected ViT model.}
    \label{tab:critical_sdc}
    \begin{tabular}{llcccc}
        \toprule
        FI & \textbf{Model} & \textbf{QKV} & \textbf{Proj} & \textbf{FC1} & \textbf{FC2} \\
        \midrule
        \multirow{3}{*}{\rotatebox[origin=c]{90}{Weight }}
        & \textbf{ViT-tiny}  & 1.83\% & 3.52\% & 2.77\% & 3.29\% \\
        & \textbf{DeiT-tiny} & 1.65\% & 3.15\% & 3.04\% & 2.93\% \\
        & \textbf{Swin-tiny} & 0.97\% & 3.25\% & 2.69\% & 2.68\% \\
        \midrule
        \multirow{3}{*}{\rotatebox[origin=c]{90}{Activation}}
        & \textbf{ViT-tiny}  & 3.39\% & 2.90\% & 2.71\% & 2.84\% \\
        & \textbf{DeiT-tiny} & 3.44\% & 2.88\% & 2.66\% & 2.86\% \\
        & \textbf{Swin-tiny} & 3.24\% & 2.67\% & 2.60\% & 1.99\% \\
        \bottomrule
    \end{tabular}
    \vspace{-5mm}
\end{table}

\begin{table*}[ht!]
\caption{Comparative results for accuracy degradation and critical SDC under FI into weights and input activations across ViT models.}
\label{tab:mitigation-fi-results}
\resizebox{\textwidth}{!}{%
\begin{tabular}{ccccccc||cccccc}
\toprule
 &
  \multicolumn{6}{c||}{\textbf{Accuracy Degradation (\%)}} &
  \multicolumn{6}{c}{\textbf{Critical SDC (\%)}} \\ \midrule  
 &
  \multicolumn{3}{c|}{\textbf{Weight FI}} &
  \multicolumn{3}{c||}{\textbf{Activation FI}} &
  \multicolumn{3}{c|}{\textbf{Weight FI}} &
  \multicolumn{3}{c}{\textbf{Activation FI}} \\ \midrule  
 &
  \multicolumn{1}{c}{ViT-Tiny} &
  \multicolumn{1}{c}{DeiT-Tiny} &
  \multicolumn{1}{c|}{Swin-Tiny} &
  \multicolumn{1}{c}{ViT-Tiny} &
  \multicolumn{1}{c}{DeiT-Tiny} &
  Swin-Tiny &
  \multicolumn{1}{c}{ViT-Tiny} &
  \multicolumn{1}{c}{DeiT-Tiny} &
  \multicolumn{1}{c|}{Swin-Tiny} &
  \multicolumn{1}{c}{ViT-Tiny} &
  \multicolumn{1}{c}{DeiT-Tiny} &
  Swin-Tiny \\ \midrule  
Unprotected &
  \multicolumn{1}{c}{$2.14\%$} &
  \multicolumn{1}{c}{$1.92\%$} &
  \multicolumn{1}{c|}{$1.92\%$} &
  \multicolumn{1}{c}{$2.22\%$} &
  \multicolumn{1}{c}{$2.12\%$} &
  $2.09\%$
   &
  \multicolumn{1}{c}{$2.85\%$} &
  \multicolumn{1}{c}{$2.69\%$} &
  \multicolumn{1}{c|}{$3.39\%$} &
  \multicolumn{1}{c}{$2.96\%$} &
  \multicolumn{1}{c}{$2.96\%$} &
  $2.62\%$ \\ \midrule  

ABFT &
  \multicolumn{1}{c}{$<\textbf{0.01\%}$} &
  \multicolumn{1}{c}{$\textbf{0.05\%}$} &
  \multicolumn{1}{c|}{$0.92\%$} &
  \multicolumn{1}{c}{-} &
  \multicolumn{1}{c}{-} &
  - &
  \multicolumn{1}{c}{$\textbf{0.47\%}$} &
  \multicolumn{1}{c}{$\textbf{0.46\%}$} &
  \multicolumn{1}{c|}{$1.3\%$} &
  \multicolumn{1}{c}{-} &
  \multicolumn{1}{c}{-} &
  - \\ \midrule
  
CheckOne &
  \multicolumn{1}{c}{$<\textbf{0.01\%}$} &
  \multicolumn{1}{c}{$\textbf{0.05\%}$} &
  \multicolumn{1}{c|}{$<\textbf{0.01\%}$} &
  \multicolumn{1}{c}{$\textbf{0.07\%}$} &
  \multicolumn{1}{c}{$\textbf{0.06\%}$} &
  $\textbf{0.10\%}$ &
  \multicolumn{1}{c}{0.70\%} &
  \multicolumn{1}{c}{0.59\%} &
  \multicolumn{1}{c|}{\textbf{0.13\%}} &
  \multicolumn{1}{c}{\textbf{0.18\%}} &
  \multicolumn{1}{c}{\textbf{0.17\%}} &
  \textbf{0.17\%} \\   
 \bottomrule  
 
\end{tabular}%
}
\vspace{-5mm}
\end{table*}

\subsection{Impact of CheckOne on Reliability and Performance}

Table~\ref{tab:mitigation-fi-results} summarizes the experimental results for unprotected ViTs, conventional ABFT, and CheckOne under both weight and activation FI campaigns. The results demonstrate that both ABFT and CheckOne significantly improve the resilience of ViTs against bit-flips in weights compared to unprotected execution. Overall, CheckOne achieves lower accuracy degradation than conventional ABFT, particularly for Swin-Tiny, while maintaining accuracy degradation consistently below $0.05\%$ across all evaluated models. Although the critical SDC rate of CheckOne is slightly higher than that of ABFT in some cases, it remains competitive. These observations indicate that CheckOne achieves reliability comparable to conventional ABFT for protecting model weights.

Unlike conventional ABFT, CheckOne can also effectively protect input activations against single bit-flips. Across all evaluated ViT models, the observed accuracy degradation under activation FI remains below $0.1\%$, while the critical SDC rate is consistently lower than $0.18\%$. Compared to unprotected ViTs, CheckOne reduces the critical SDC rate by up to $26\times$ under weight FI and up to $17\times$ under activation FI. Overall, these results confirm that CheckOne provides efficient and consistent protection for both weights and activations against soft errors in transformer-based architectures.



Beyond its reliability advantages, CheckOne also delivers substantially higher performance efficiency compared to conventional ABFT, primarily by eliminating the costly recomputation of checksums during inference. Performance evaluations conducted on an NVIDIA A100 GPU over $100$ inference runs using ViT-Tiny, DeiT-Tiny, and Swin-Tiny demonstrate that CheckOne achieves execution speedups of $4.67\times$, $4.72\times$, and $2.02\times$, respectively, relative to the conventional ABFT implementation. Overall, CheckOne provides an average performance improvement of $3.8\times$ while maintaining a comparable level of resilience against soft errors. Note that the additional memory overhead introduced by CheckOne is negligible for storing precomputed values, accounting for less than $0.3\%$ of the total memory footprint of the evaluated ViT models. The evaluation results indicate that CheckOne is a significantly more practical and efficient fault tolerance solution for transformer-based architectures.


%% file: sections/5-conclusions.tex
\section{Conclusions} \label{sec:conclusion}

This work presents CheckOne, an innovative, cost-effective method for protecting ViTs against faults in on-chip memories during inference. In CheckOne, the inputs to a matrix multiplication are protected by a single additional vector of $1$, implicitly producing their summations at the output matrix. Utilizing the pre-stored detection values, CheckOne conducts on-the-fly error detection and mitigation. Results indicate that CheckOne effectively improves the reliability of ViTs by reducing critical faults by up to $26\times$ while being $3.8\times$ faster than conventional ABFT, on average. 